\documentstyle[epsfig]{report}
\makeatletter
\renewenvironment{thebibliography}[1]
     {\section*{\bibname
        \@mkboth{\MakeUppercase\bibname}{\MakeUppercase\bibname}}%
      \list{\@biblabel{\@arabic\c@enumiv}}%
           {\settowidth\labelwidth{\@biblabel{#1}}%
            \leftmargin\labelwidth
            \advance\leftmargin\labelsep
            \@openbib@code
            \usecounter{enumiv}%
            \let\p@enumiv\@empty
            \renewcommand\theenumiv{\@arabic\c@enumiv}}%
      \sloppy
      \clubpenalty4000
      \@clubpenalty \clubpenalty
      \widowpenalty4000%
      \sfcode`\.\@m}
     {\def\@noitemerr
       {\@latex@warning{Empty `thebibliography' environment}}%
      \endlist}
\makeatother
\renewcommand{\author}[1]{\subsubsection*{#1}}
\newcommand{\address}[1]{\subsubsection*{\it #1}}

\begin{document}

\chapter*{Oscillations in solar--type stars tidally induced by 
orbiting planets}

\author{C. Terquem$^1$, J.C.B. Papaloizou$^2$, R.P. Nelson$^2$ and
D.N.C. Lin$^1$}

\address{$^1$UCO / Lick Observatory, University of California,
Santa-Cruz, USA \\ $^2$Astronomy Unit, Queen Mary~\& Westfield
College, London, UK}

\section{Introduction}

We examine the effect of dynamical tides raised by a companion on a
solar--type star. In these binaries, gravity or g~mode oscillations
are excited by the companion in the radiative region beneath the
convective envelope of the star. They become evanescent in the
convection zone.

This is of particular interest in connection with the newly discovered
planets, some of which are found to orbit around solar--type stars
with a period comparable to that of the high order g~modes of the
star. One such example is 51~Pegasi (\cite{Mayor1}; \cite{Marcy1}).

Here, we determine the magnitude of the perturbed velocity induced by
the tides at the stellar surface. We show that, in the case of
51~Pegasi, this velocity is too small to be observed. This result is
insensitive to the magnitude of the stellar turbulent viscosity
assumed and is not affected by the possibility of resonance, which
occurs when the frequency of the tidal disturbance is close to that of
some normal mode of the star. We also discuss the orbital evolution
and synchronization timescales associated with the tidal interaction,
a detailed calculation of which  will be presented elsewhere.

\section{Tidal response to a companion in circular orbit}

In the case of 51~Pegasi, observations indicate that the rotational
period of the star is between 30 and 40 days (\cite{Baliunas}). Since
this is much larger than the orbital period (4.23 days), we neglect
the rotational angular velocity of the primary compared to the orbital
frequency. Quadrupolar tidal forcing thus occurs through potential
perturbations with periods which are half the orbital period.

The tides raised on the star are damped by turbulent friction
(\cite{Zahn1}) in the convective envelope, and by non--adiabaticity
arising from heat transport in the radiative interior
(\cite{Zahn2}). For the periods of interest, these dissipative
mechanisms are predominant away from and close to resonance
respectively. We then calculate the tidal response in
resonance taking radiative damping alone into account. Away from
resonance, we should in principle include turbulent dissipation to
calculate the tide. However, dissipation in the convection zone of
solar--type stars is weak enough so that the tidal response away from
resonance is well approximated by assuming adiabaticity. 

The linearized momentum and mass equations governing the response of
the non--rotating star to the perturbing potential $\Psi_T$ may be
written (\cite{Unno})

\begin{equation}
\frac{\partial^2 \mbox{\boldmath $\xi$}}{\partial t^2} = -
\frac{1}{\rho} \mbox{\boldmath $\nabla$} P' + \frac{\rho'}{\rho^2}
\mbox{\boldmath $\nabla$} P - \mbox{\boldmath $\nabla$} \Psi_T ,
\label{ct:momentum} 
\end{equation}

\begin{equation}
\rho' = - \mbox{\boldmath $\nabla$} \cdot \left( \rho \mbox{\boldmath
$\xi$} \right) ,
\label{ct:mass} 
\end{equation}

\noindent where $P$ is the pressure, $\rho$ is the density,
$\mbox{\boldmath $\xi$}$ is the Lagrangian displacement vector and the
primed quantities are Eulerian perturbations.

In the adiabatic approximation, the energy equation is $\Delta S =0$,
where $S$ is the entropy per unit mass  and $\Delta$ denotes the Lagrangian
perturbation. When non--adiabaticity in the radiative core is taken
into account, this equation takes on the form: $\rho T \partial \left(
\Delta S \right)/\partial t = - \mbox{\boldmath $\nabla$} \cdot {\bf
F}'$, where $T$ is the temperature and ${\bf F}'$ is the perturbed
radiative flux. We suppose that close to resonance, the response
behaves exactly like a free g~mode with very large radial wavenumber
so that WKB theory can be used together with the local dispersion
relation to estimate $\mbox{\boldmath $\nabla$} \cdot {\bf F}'$ in
this last equation. In addition, $\Delta S$ is related to $\Delta P$
and $\Delta \rho$ through a standard thermodynamic relation.

Here, only the companion's dominant tidal term is considered in the
perturbing potential (\cite{Cowling}). For a system with a circular
orbit, this is given in spherical polar coordinates
$(r,\theta,\varphi)$ by the real part of

\begin{equation} 
\Psi_T \left( r, \theta, \varphi, t \right) = f r^2 \, P_n^{|m|}
(\cos\theta) \,\, e^{i m \left( \varphi- \omega t \right)}
\label{ct:eqpot} 
\end{equation}

\noindent with $n=m=2$, $P^{|m|}_n$ being the associated Legendre
polynomial. Here $\omega$ is the orbital angular velocity, $D$ is the
orbital separation, and $f=-GM_p/4D^3,$ with $M_p$ being the mass of
the companion. The Lagrangian displacement can then be sought in
the form:

\begin{equation} 
\mbox{\boldmath $\xi$} = \left[ \xi_r(r), \xi_h(r) {\partial \over
\partial \theta}, \xi_h(r) {\partial \over {\sin \theta \partial
\varphi} } \right] P_n^{|m|} (\cos\theta) \,\, e^{i m \left( \varphi-
\omega t \right)} .
\label{ct:xi}
\end{equation} 

\noindent We can eliminate $P'$ and $\rho'$ using the non--radial
momentum and energy equations. The radial momentum and mass equations
then reduce to a pair of ordinary differential equations for $\xi_r$
and $\xi_h$:

\begin{eqnarray}
\frac{d \xi_r}{d r} & = & \left( - \frac{2}{r} + \bar{A} - \frac{dln
\rho}{dr} \right) \xi_r + \left[ - \frac{m^2 \omega^2 r \rho}{\Gamma_1
P} + \frac{n(n+1)}{r} \right] \xi_h + \frac{f r^2 \rho}{\Gamma_1 P},
\label{ct:ODE1} \\ 
\frac{d \xi_h}{dr} & = & \frac{1}{r} \left( 1 - \frac{\bar{A}P}{m^2
\omega^2 \rho} \frac{dlnP}{dr} \right) \xi_r - \left( A + \frac{1}{r}
\right) \xi_h + \frac{Afr}{m^2 \omega^2} ,
\label{ct:ODE2}
\end{eqnarray}

\noindent where $A=dln\rho/dr - (dlnP/dr)/\Gamma_1$, $\Gamma_1$ being
the adiabatic exponent. We have defined $\bar{A}$ such that
$\bar{A}=A$ in the adiabatic approximation, and, when radiative
damping is taken into account, $\bar{A}=A/(1+i\epsilon)$ with
$\epsilon=16 a c T^4 N^2/(5 m^3 \omega^3 \kappa \rho P r^2)$ in the
radiative core and $\epsilon=0$ elsewhere. Here $a$ is the
Stefan--Boltzmann radiation constant, $c$ is the velocity of light,
$\kappa$ is the opacity and $N^2=-Ag$ is the square of the
Brunt-V\"ais\"al\"a frequency, $g$ being the acceleration due to
gravity.

\noindent The solution of this system requires two boundary
conditions. At the surface of the star we take a free boundary. At
$r=0$, where equations~(\ref{ct:ODE1}) and~(\ref{ct:ODE2}) have a
regular singularity, the boundary condition is that the solutions be
regular.

\section{Numerical results: tidal response and velocity at the surface 
of the star}

The calculations presented in the previous section are applied to a
standard solar model (\cite{Christensen}). We solve numerically the
differential equations~(\ref{ct:ODE1}) and~(\ref{ct:ODE2}) using a
shooting method to an intermediate fitting point. We define $x \equiv
r/R_c,$ where $R_c$ is the outer radius of the convective
envelope. With this notation, the equations are integrated from
$x_{in}=10^{-6}$ to $x_{out}=1.00071256.$ The radiative core extends
from $x=0$ to $x \simeq 0.7$.

\begin{figure}
\epsfxsize=15.0 cm
     \epsfysize=15.0 cm
      {\epsffile{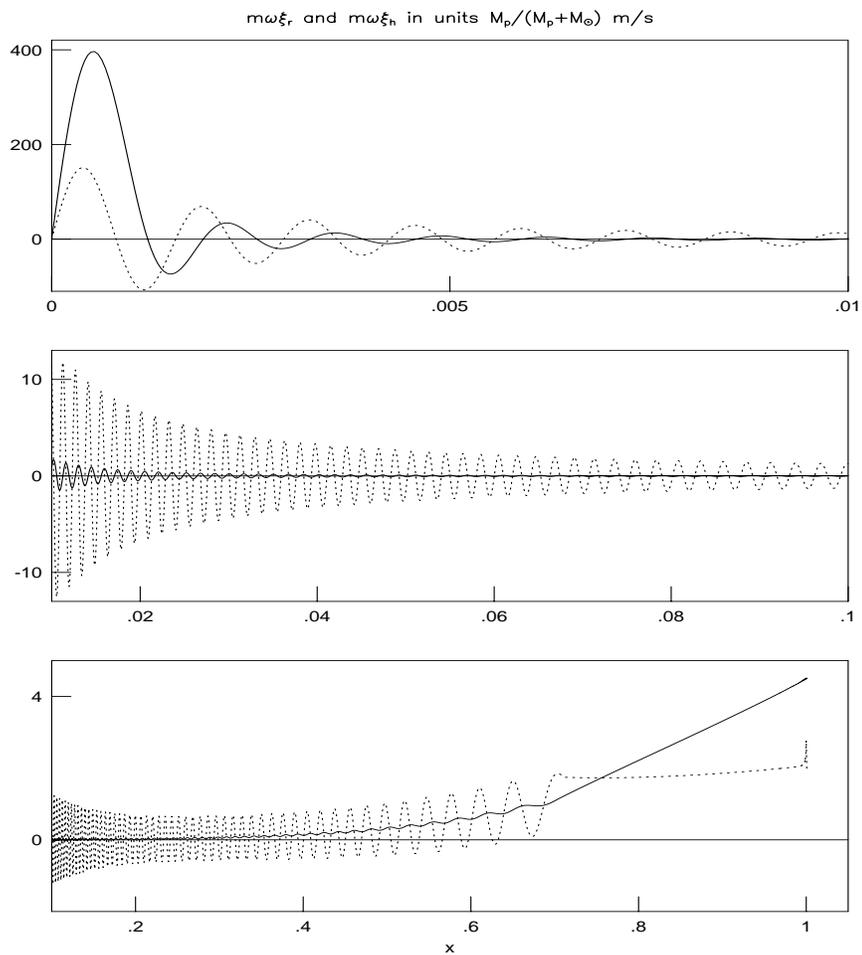}}
\caption[]{Real part of $m \omega \xi_r$ (solid lines) and $m \omega
\xi_h$ (dotted lines) in units $M_p/(M_p+M_{\odot})$~$m/s$ versus $x$
for $x_{in} \le x \le 0.01$ (top panel), $0.01 \le x \le 0.1$ (middle
panel) and $0.1 \le x \le x_{out}$ (bottom panel), and for
$P_o=4.23$~$d$.}
\label{ct:fig}
\end{figure}

In Figure~\ref{ct:fig}, we plot the spatial distribution of the real
parts of $m \omega \xi_r$ and $m \omega \xi_h$ for an orbital period
of $P_o=4.23$~$d$ away from resonance and in the adiabatic
approximation. Away from resonance, the magnitude of the imaginary
parts of these quantities is much smaller than that of their real
parts (which is why the adiabatic approximation can be
used). Therefore, they represent typical values of the radial and
horizontal velocities, the maximum values being three and six times
larger respectively. Since these quantities depend on the perturbing
mass through the ratio $M_p/(M_p+M_{\odot})$, they have been
represented in units of this factor.

We see from Figure~\ref{ct:fig} that a companion orbiting around the
star with a period $P_o=4.23$~$d$ induces a radial velocity at the
stellar surface the maximum of which is between $10^{-2}$ and 6~$m/s$
for $M_p$ between $10^{-3}$ and 1~$M_{\odot}$. This is at least one
order of magnitude smaller than the observed velocity. The period of
the tidal oscillation corresponding to this orbital period is
2.115~$d$. For the oscillation to have a period of $4.23$~$d$, the
orbital period would have to be 8.46~$d$. The maximum perturbed radial
velocity at the surface of the star induced by the companion would
then be between $2\times10^{-3}$ and 1~$m/s$ for a perturbing mass
between $10^{-3}$ and 1~$M_{\odot}$. These velocities are at least
about 50 times smaller than the observed ones. {\it These numbers do
not depend on the magnitude of the turbulent viscosity assumed}. We
have also checked that, because of evanescence in the convective
envelope, {\it they are not affected by the possibility of resonance}.

\section{Discussion and Conclusion}

The planetary companion interpretation has been questioned recently by
the reported 4.23~$d$ modulation in the line profile of 51~Pegasi
(\cite{Gray1}), and the possibility that this modulation may be due to
g~mode oscillations has been considered (\cite{Gray2}). We note that,
according to our results, such a modulation could not be due to g~mode
oscillations tidally driven by a companion. 

From the calculations presented above, it is also possible to compute
the various timescales associated with the tidal interaction. For
$P_o=4.23$~$d$ and $M_p=10^{-3}$~$M_{\odot}$, it is found that if the
turbulent viscosity of the star is calibrated such that the
calculations can account for the observed circularization rates of
main sequence solar--type binaries, the tidal orbital evolution,
circularization, stellar spin up and convective envelope spin up
timescales are respectively 143, 24, 131 and 18~$Gyr$. All of these
timescales are long compared with the inferred age of 51~Pegasi
(\cite{Edvardsson}). If the companion is a low-mass star of $0.1
M_{\odot}$, as has been recently suggested, these numbers drop to 1.7,
0.25, 0.016 and 0.0022~$Gyr$ respectively. We then expect the primary
star to be synchronized with the orbit, in which case exchange of
angular momentum is no longer taking place.  Synchronization is
actually expected if the mass of the companion is larger than about 10
Jupiter masses. We note that since the orbital decay timescale is
larger than the synchronization one, tidal interaction stops before
the companion has plunged into the central star.

If a simple estimate of the turbulent viscosity based on the usual  mixing
length theory is used, all these timescales have to multiplied by
$\sim 50$. In that case, synchronization is expected if the mass of
the companion is larger than about 70 Jupiter masses. \\

This work is supported by PPARC through grant GR/H/09454, by NSF and
NASA through grants AST~93--15578 and NAG~5--4277, and by the Center
for Star Formation Studies at NASA/Ames Research Center and the
University of California at Berkeley and Santa-Cruz.

\end{document}